\documentclass[preprint,showpacs,aps,preprintnumbers,amsmath,amssymb,
nofootinbib]{revtex4}
\usepackage{epsfig}
\begin{document}

\begin{flushright}
%\preprint{hep-ph/xxyyzz}
%\today\\
%UH-0528
\end{flushright}

%%%%%%%%%%%%%%%%%%%%%%%%%%%%%%%%%%%%%%%%%%%%%%%%%%%%%%%%%%%%%%%%%%%%%%%%

\newcommand{\be}{\begin{equation}}
\newcommand{\ee}{\end{equation}}
\newcommand{\bea}{\begin{eqnarray}}
\newcommand{\eea}{\end{eqnarray}}
\newcommand{\bers}{\begin{eqnarray*}}
\newcommand{\eers}{\end{eqnarray*}}
\newcommand{\nn}{\nonumber}

\newcommand{\Apa}{A_\parallel}
\newcommand{\Ape}{A_\perp}
\newcommand{\dab}{\delta_{ab}}
\newcommand{\dca}{\delta_{ca}}
\newcommand{\dcb}{\delta_{cb}}
\newcommand{\da}{\delta_{a}}
\newcommand{\db}{\delta_{b}}
\newcommand{\dc}{\delta_{c}}
\newcommand{\pab}{\varphi_{ab}}
\newcommand{\pcb}{\varphi_{cb}}
\newcommand{\pca}{\varphi_{ca}}
\newcommand{\pa}{\varphi_{a}}
\newcommand{\pb}{\varphi_{b}}
\newcommand{\pc}{\varphi_{c}}
\def\ss{\scriptstyle}

\def\tr{\mbox{tr}\,}
\def\Tr{\mbox{Tr}\,}
\def\dag{^\dagger}
\def\res{\mbox{Res}}
\def\re{\mbox{Re}\,}
\def\b{\bigskip}
\def\s{\smallskip}
\def\l{\hspace*{0.05cm}}
\def\esp{\hspace*{1cm}}

\def\barp{{\raise.35ex\hbox
{${\sss (}$}}---{\raise.35ex\hbox{${\sss )}$}}}
\def\barpd{{\raise.35ex\hbox
{${\sss (}$}}--{\raise.35ex\hbox{${\sss )}$}}}
\def\dbbarp{\hbox{$B^{0}$\kern-1.2em\raise1.5ex\hbox{\barpd}}}
\def\kbarp{\hbox{$K^{*0}$\kern-1.6em\raise1.5ex\hbox{\barpd}}}

\def\sss{\scriptscriptstyle}
\def\barp{{\raise.35ex\hbox
{${\sss (}$}}--{\raise.35ex\hbox{${\sss )}$}}}
\def\bark{{\raise.35ex\hbox
{${\sss (}$}}--{\raise.35ex\hbox{${\sss )}$}}}
\def\barpd{{\raise.35ex\hbox
{${\sss (}$}}--{\raise.35ex\hbox{${\sss )}$}}}
\def\dbarp{\hbox{$D^{*0}$\kern-1.35 em\raise1.5ex\hbox{\barpd}}}
\def\dbark{\hbox{$K^{*0}$\kern-1.35 em\raise1.5ex\hbox{\barpd}}}
\def\bdbarp{\hbox{$B_d^0$\kern-1.1 em\raise1.4ex\hbox{\barp}}}
\def\p0{\hbox{$D^{0}$\kern-1.25 em\raise1.5ex\hbox{\barpd}}}
\def\rp1{\hbox{$D^{*0}$\kern-1.55 em\raise1.5ex\hbox{\barpd}}}
\def\arp1{\hbox{$B_d$\kern-1.1 em\raise1.4ex\hbox{\barp}}}
\def\drp2{\hbox{$K^{*0}$\kern-1.6 em\raise1.5ex\hbox{\barpd}}}
%%%%%%%%%%%%%%%%%%%%%%%%%%%%%%%%%%%%%%%%%%%%%%%%%%%%%%%%%%%%%%%%%%%%%%%
\title{ Weak phase information from the color suppressed
$\bdbarp~ \to \dbarp~~ \dbark $ modes}
\author{A. K. Giri$^1$, B. Mawlong$^2$ and R. Mohanta$^2$}
\affiliation{$^1$Department of Physics,\\ Punjabi University,
Patiala - 147002, India\\
$^2$School of Physics, University of Hyderabad,
Hyderabad - 500046, India\\}

%\vspace*{0.2 truein}

\begin{abstract}
The decay channels $\arp1~ \to \p0~\drp2~,~\rp1~~ \drp2 ~~$ are
investigated for
extracting weak $CKM$ phase information. These channels are
described by color-suppressed tree diagrams only and are free from
penguin contributions. The branching ratios for these channels are
found to be $\sim \cal O $ $(10^{-5} - 10^{-6} )$ which can be
measured at the currently running $B$ factories. The method
presented here may be well-suited to determine the
CKM angle $\gamma$.
\end{abstract}

\maketitle

After the discovery of $CP$ violation in $K$ systems in 1964, there
was  a lot of enthusiasm to look for $CP$ violation in other systems
as well. $CP$ violation is found to be rather small in $K$ systems,
however, theoretical predictions have suggested that the $B$ system
may be an ideal place for detecting sufficiently large $CP$
violating effects. In fact large CP violation in $B$ system has been
observed in the currently running dedicated $B$ factories such as
BaBar at SLAC and BELLE at KEK. Over the years, a lot of work have
been done in the area of $CP$ violation with many facilities
collecting data on $B$ events. In the standard model (SM), $CP$
violation arises from the non-zero weak phase in the complex
Cabibbo-Kobayashi-Maskawa (CKM) matrix which is responsible for the
charged-current  weak interaction and is characterized by the
so-called unitarity triangle. It is, therefore, imperative that the
three  angles of this triangle, termed as $\alpha, \beta$ and
$\gamma$, be measured independently to get decisive information on
the origin of $CP$ violation.

Usually, these angles are extracted from $CP$ violating rate
asymmetries in $B$ decays. The angle $\beta$ (or $\sin 2 \beta$) has
been cleanly determined from the measurement of the time dependent
$CP$ asymmetry in the golden decay mode $B_{d}^0  \to J / \psi
K_{S}$ \cite{expt}. The angle $\alpha$  can be measured using the $CP$
asymmetries in $B_{d}^0 \to \pi^+ \pi^-$ \cite{ref33},
but due to the existence of
penguin diagrams there are theoretical hadronic uncertainties which
are very difficult to quantify. The last angle which is hoped to be
determined cleanly is $\gamma$ = arg($ - V_{ud} V_{ub}^* / V_{cd}
V_{cb}^* $). There have been many attempts, suggestions and
discussions to measure this angle as cleanly as possible. The
Gronau, London and Wyler (GLW) \cite{GLW} method, which is usually
termed as the golden method to extract $\gamma$, employs the $B \to
D K $ modes to extract this angle. In the GLW method the
interference of two amplitudes ($b \to c \bar u s$ and $b \to u \bar
c s$) is being used which endows us to determine the angle $\gamma$
by measuring the decay rates for $B \to D^0 K, B \to \bar D^0 K$ and
$B \to D_{CP}^0 K$ and their corresponding $CP$ conjugate modes.
However, it is experimentally difficult to measure the mode $B^- \to
\bar D^0 K^-$. As pointed out in earlier studies, the reason is that
the final $\bar D^0$ meson should be identified using $\bar D^0 \to
K^+ \pi^-$, but it is difficult to distinguish it from the doubly
Cabibbo suppressed $D^0 \to K^+ \pi^-$.

There exist many studies in the literature \cite{ads} to help
overcome the difficulty and to provide improved ways to determine
the angle $\gamma$. It is highly desirable to have independent
measurement of the angle $\gamma$ (or otherwise the angle $\alpha$),
at least to the precision of the angle $\beta$ as of today, to
understand better the CKM mechanism of CP violation under the
framework of the SM. But so far we have not been able to succeed in
this effort. Given the various methods, and wide range of options
available, the measurement of the angle $\gamma$ seems to be a
better option. This is currently being done and will also be taken
up in the second generation experiments. There is also another
parameter, namely, $2\beta +\gamma$ which is also discussed in the
literature \cite{dunietz} to be measured. Since $\beta$ is well
measured by now, therefore, the measurement of $2\beta +\gamma$ will
be very much useful for the clean determination of $\gamma$. It
should be noted here that we should measure the angle $\gamma$ in
all possible way (and as cleanly as possible) to independently
verify the measurements, improve the statistics and to help resolve
discrete ambiguities. To this end, we intend to  present here
another important and simple way to extract the weak phase
$\gamma/(2\beta +\gamma)$ from the decay modes $\arp1~ \to
\p0~\drp2~,~\rp1~~ \drp2 ~~$.

In this paper, we consider the color-suppressed decay modes $\arp1~
\to \p0~(\rp1~~) ~\drp2~$,  to extract the CKM phase information.
Several studies have been carried out using these decay modes
\cite{ko, akiba, sp} for the extraction of the angle $\gamma$. We,
in this investigation, present  another alternative method, which is
very clean and simple to extract the weak phase $\tan^2 \gamma~
(\tan^2(2\beta+\gamma))$. Since, as emphasized before, one needs to
have as many clean methods as possible to improve the sensitivity
and to resolve the discrete ambiguities the present method will be
very much helpful in that direction.

 First let us consider the decay channels $B^0 \to D^0 K^{*0}$
 and $B^0 \to \bar D^{0} K^{* 0}$. It has been shown in Ref.
 \cite{akiba} that the CKM angle $\gamma$ can be determined by
 measuring the following six decay rates: $B^0 \to D^0
 K^{*0}$
 and $B^0 \to \bar D^{0} K^{* 0}$, $B \to D_{\rm CP} K^{*0}$ (where
 $D_{\rm CP}=(D^0 + \bar D^0)/\sqrt 2$, is the CP even eigenstate
 of neutral $D$ meson) and the corresponding conjugate processes.
The $D^0 (\bar D^0)$ meson is considered to decay subsequently to
the flavor state  $K^+ \pi^-$ for which the ratio of the two
amplitudes is found to be very tiny i.e., $r_D=|A(B^0
 \to K^+ \pi^-)/A(\bar D^0 \to K^+ \pi^-)|=0.06 \pm 0.003$ \cite{pdg}.
Here we will show that if we consider the the decay of $D$ meson to
non-CP final state  i.e., $K^{*+} K^-$ for which $r_D \sim {\cal
O}(1)$, then it is possible to extract the CKM angle $\gamma$ by
measuring only the four decay rates. This method is very promising
because the experimental branching ratio for the  process $B^0 \to
\bar D^0 K^{* 0}$ has already been known with value $Br( B^0 \to
\bar D^0 K^{* 0}=(5.3 \pm 0.8) \times 10^{-5}$ and for the $(B^0 \to
D^0 K^{*0})$ process we have the  upper limit as $Br(B^0 \to D^0
K^{*0}) < 1.8 \times 10^{-5}$ \cite{pdg}. The advantage of using the
non-CP eigenstate  has been discussed in \cite{yg}, in connection
with the charged $B$ decays $B^\pm \to K^\pm D^0 (\bar D^0)$, which
renders the corresponding interfering amplitudes to be of same
order.

Now let us denote the amplitudes for these processes as \be A_B=
{\cal A}(\bar B^0 \to D^0 \bar K^{*0})\;,~~~~\bar A_B={\cal A}(\bar
B^0 \to \bar D^0 \bar K^{*0})\;, \ee and their ratio as \be
\frac{\bar A_B}{A_B}=r_B e^{i(\delta_B -\gamma)}\;, ~~~{\rm
with}~~~r_B=\left |\frac{\bar A_B}{A_B} \right |~~~ {\rm and}~~~
{\rm arg} \left (\bar A_B/A_B \right )=\delta_B-\gamma\;, \ee where
$\delta_B$ and $(-\gamma)$ are the relative strong and weak phases
between the two amplitudes. The ratio of the corresponding CP
conjugate processes are obtained by changing the sign of the weak
phase $\gamma$. One can then obtain a rough estimate of $r_B$ from
dimensional analysis, i.e., \be r_B=\left |\frac{V_{ub}
V_{cs}^*}{V_{cb} V_{us}^*} \right |  \approx 0.4\;.\label{rb} \ee

Now we consider that both $D^0$ and $\bar D^0$ will decay into the
common non-CP final state $(K^{*+} K^-)$.
 Denoting the  $D^0$ decay amplitudes as \be A_D= {\cal A}(D^0
\to K^{*+} K^-)\;,~~~~\bar A_D={\cal A}( \bar D^0 \to K^{*+} K^-)\;,
\ee one can write their ratio  \be \frac{\bar A_D}{A_D}=r_D
e^{i\delta_D}\;, ~~~{\rm with}~~~r_D=\left |\frac{\bar A_D}{A_D}
\right |\;, \ee where $\delta_D$ is the relative strong phase
between them. It is interesting to note that the parameters $r_D$
and $\delta_D$ have recently been measured by CLEO collaboration
\cite{cleo}, with values $r_D=0.52\pm0.05\pm0.04$ and
$\delta_D=332^\circ \pm 8^\circ \pm 11^\circ $.

With these definitions the four amplitudes are given as \bea {\cal
A}_1(\bar B_d^0 \to  (K^{*+} K^-)_{\rm D}\bar K^{*0} ) &=& |A_B
A_D|\Big[1+r_B r_D e^{i(\delta_B+
\delta_D-\gamma)}\Big]\;,\nn\\
{\cal A}_2(\bar B_d^0 \to   (K^{*-} K^+)_{\rm D}\bar K^{*0}) &=&
|A_B A_D| e^{i \delta_D}\Big[r_D+ r_B e^{i(\delta_B-
\delta_D-\gamma)}\Big]\;,\nn\\
{\cal A}_3( B_d^0 \to   (K^{*-} K^+)_{\rm D}  K^{*0}) &=& |A_B
A_D|\Big[1+r_B r_D e^{i(\delta_B+
\delta_D+\gamma)}\Big]\;,\nn\\
{\cal A}_4( B_d^0 \to  (K^{*+} K^-)_{\rm D}  K^{* 0}) &=& |A_B
A_D|e^{i \delta_D}\Big[r_D+ r_B e^{i(\delta_B-
\delta_D+\gamma)}\Big]\;.\label{eq} \eea From these amplitudes one
can obtain the four observables ($R_1, \cdots, R_4$), with the
definition \be R_i = \left |{\cal A}_i(\arp1~ \to (K^{*\pm} K^\mp)_D
~ ~\drp2~~ )/(A_B A_D)\right |^2\;.\label{eq8} \ee We can now write
$R_1 = 1+r_B^2 r_D^2 + 2 r_B r_D \cos (\delta_B+\delta_D -\gamma)$
and similarly $R_2$, $R_3$ and $R_4$. Thus, we get four observables
and three unknowns, namely, $r_B$, $\delta_B$ and $\gamma$. Hence
$\gamma$ can in principle be determined from these four observables.

Assuming that the amplitudes $|A_B|$ and $|A_D|$ are known (so also
$r_B$, which is expected to be $\sim{\cal O}$(0.4)), we obtain
 an analytical expression for $\gamma$ as \be
\tan^2 \gamma=\frac {(R_1-R_3)^2 -(R_2-R_4)^2 }{
[R_2+R_4-2(r_B^2+r_D^2)]^2-[R_1+R_3-2(1+r_B^2 r_D^2)]^2
}\;.\label{eq1} \ee Thus the measurement of the four observables
$R_{1, \cdots, 4}$ can be used to extract cleanly the CKM angle
$\gamma $.

Next, we consider the decay channels $B_d^0 \to D^{*0} K^{*0}$, $\bar
D^{*0} K^{*0}$ and $\bar B_d^0 \to D^{*0} \bar K^{*0}$, $\bar D^{*0}
\bar K^{*0}$ with two vector mesons in the final state.
Considering the decay of a $B$ meson into two vector mesons
$V_1$ and $V_2$, which subsequently decay into pseudoscalar mesons
i.e.,  $V_1 \to P_1 P_1'$ and $V_2 \to P_2 P_2'$, one can
write the normalized differential
angular distribution as \cite{chiang1},
\begin{eqnarray}
&&\frac{1}{\Gamma} \frac{d^3 \Gamma}{d \cos \theta_1 \l d \cos
\theta_{2} \l d \psi}  =   \frac{9}{8 \pi \Gamma } \;\Bigg\{ L_1 \l
\cos^2 \theta_{1} \l \cos^2 \theta_{2} \l + \l \frac{L_2}{2} \l
\sin^2 \theta_{1} \l
\sin^2 \theta_{2} \l \cos^2 \psi \nonumber \\
&& \hspace{2.5 cm}  + \l \frac{L_3}{2} \l \sin^2 \theta_{1} \l
\sin^2 \theta_{2} \l \sin^2 \psi \l + \l \frac{L_4}{2 \sqrt{2}} \l
\sin 2 \theta_{1} \l \sin 2 \theta_{2}
\l \cos \psi \nonumber \\
&& \hspace{2.5 cm}  -  \l \frac{L_5}{2 \sqrt{2}} \l \sin 2
\theta_{1} \l \sin 2 \theta_{2} \l \sin \psi \l - \l \frac{L_6}{2}
\l \sin^2 \theta_{1} \l \sin^2 \theta_{2} \l \sin 2 \psi \Bigg \}\,,
\label{dg}
\end{eqnarray}
where $\theta_{1}$ ($\theta_2$) is the angle between the
three-momentum of $P_1$ ($P_2$) in the $V_1$ ($V_2$) rest frame and
the three-momentum of $V_1$ ($V_2$) in the $B$ rest frame, and $\psi$
is the angle between the normals to the planes
defined by $P_1 P_1'$ and $P_2 P_2'$, in the $B$
rest frame. The coefficients $L_i$ can be expressed in terms of
three independent amplitudes, $A_0$, $A_\|$ and $A_\bot$, which
correspond to the different polarization states of the vector mesons
 as
\begin{eqnarray}
L_1 = |A_0|^2\;, \hspace{2.cm} & L_4 = {\rm Re} [A_{\|} A^*_0]\;,
\nonumber \\ L_2 = {|A_{\|}|}^2\;, \hspace{2.cm} & L_5 = {\rm Im}
[A_\bot A^*_0]\;, \nonumber \\  L_3 =
|A_{\bot}|^2\;, \hspace{2.cm} & L_6 = {\rm Im} [A_{\bot}
A^*_{\|}]\;. \label{K}
\end{eqnarray}
In the above $A_0$, $\Apa$, and $\Ape$ are complex amplitudes of the
three helicity states in the transversity basis. These observables
can be efficiently extracted from the angular distribution (\ref{dg})
using the appropriate weight functions as discussed in Ref. \cite{dighe}.

The decay mode $B\to V_1V_2$ can also be described in the helicity
basis, where the amplitude for the helicity matrix element can be
parameterized as~\cite{kramer}
\begin{eqnarray}
H_{\lambda}&=& \langle V_1 (\lambda)V_2(\lambda)|{\cal H}_{eff} |B^0
\rangle\nonumber\\
&=& \varepsilon_{1 \mu}^* (\lambda) \l \varepsilon_{2
\nu}^* (\lambda) \left [ a g^{\mu \nu} + \frac{b}{m_1 m_2} p^{\mu}
p^{\nu} + \frac{i c}{m_1 m_2} \epsilon^{\mu \nu \alpha \beta} p_{1
\alpha} p_{\beta} \right ]\;,
\label{hlam}
\end{eqnarray}
where $p$ is the $B$ meson momentum, $\lambda =0, \pm 1$ are
the helicity of both the
vector mesons and $m_i$, $p_i$ and $\varepsilon_i$
($i=1,2$) denote their
masses, momenta and polarization vectors respectively.
Furthermore, the three invariant amplitudes $a$, $b$, and $c$ are related to
the helicity amplitudes by
\begin{equation}
H_{\pm 1} = a \pm c \l \sqrt{x^2 - 1}\;,
\esp
H_0 = - a x - b \l (x^2 - 1)\;,
\label{a1}
\end{equation}
where $x =(p_1 \cdot p_2)/m_1 m_2 = (m_B^2 - m^2_1 - m^2_2)/(2 m_1 m_2)$.

The corresponding decay rate using the helicity basis amplitudes can
be given as
\begin{equation}
\Gamma = \frac{p_{cm}}{8 \pi m_B^2} \biggr( |H_0|^2+|H_{+1}|^2 +|H_{-1}|^2
\biggr)\;,
\end{equation}
where $p_{cm}$ is the magnitude center-of-mass momentum of the outgoing vector
particles.

The  amplitudes in  transversity  and helicity basis are related to each other
through the following relations
\begin{eqnarray}
A_{\bot} \l = \l \frac{H_{+1} - H_{-1}}{\sqrt{2}}, \esp
A_{\|} \l = \l \frac{H_{+1} + H_{-1}}{\sqrt{2}}, \esp A_0 \l =
\l H_0 \label{cb}.
\end{eqnarray}

The corresponding helicity amplitudes $\bar H_{\lambda}$ for the
complex conjugate decay process $\bar B \to \bar V_1 \bar V_2$ have
the same decomposition with $a \to \bar a, b \to \bar b$ and $c \to
-\bar c$.  The amplitudes $\bar a, \bar b$ and $\bar c$ can be
obtained from $a, b$ and $c$ by changing the sign of the weak
phases.

In order to study the feasibility of this method, first we would
like to estimate the branching ratios of the above mentioned decay
modes. Only, the experimental upper limits for these modes are known
so far  i.e., $Br (B^0 \to \bar D^{*0} K^*) < 6.9 \times 10^{-5}$
and $Br (B^0 \to D^{*0} K^*) < 4.0 \times 10^{-5}$ \cite{pdg}. We
expect that these modes will be well measured soon  in the currently
running asymmetric $B$ factories or in the upcoming LHCb experiment.

In the SM, these decays proceed
through color suppressed tree diagrams only and are free
from  penguin contributions. The decay  $B^0 \to D^{*0} K^{*0}$ arises
from the quark
level transition $\bar b \to \bar u c \bar s$ and the process
$\bar B^0 \to D^{*0} \bar K^{*0}$ arises from $b \to c \bar u s$.
To evaluate the hadronic matrix element
$\langle O_i \rangle \equiv
\langle D^{*0} \drp2~~ | O_i | \arp1~~ \rangle$, the factorization
approximation has been used. Thus, in this approach, we obtain the
factorized amplitude for the $ B_0 \to D^{* 0} K^{*0}$  modes as
\begin{eqnarray}
H &=& \frac{G_{F}}{\sqrt{2}} \lambda_u^*~ a_2
 \langle K^{*0} (\varepsilon_1, p_1)| {(\bar s b)}_{V-A}|B_{d}^{0}
(p)\rangle \langle D^{* 0}(\varepsilon_2, p_2)
|{(\bar u c)}_{V-A}|0 \rangle\nonumber\\
&=&\frac{G_{F}}{\sqrt{2}} \lambda_u^*~ a_2 ~ i f_{D^{*0}} m_{D^{*0}}
\biggr[(m_{B^{0}}+m_{K^{*0}}) A_1^{B K^*} (m_{D^{*0}}^2)
(\varepsilon_{1}^{*} \cdot \varepsilon_{2}^{*})\nonumber \\
 &-&   \frac{2 A_{2}^{B K^*}
(m_{D^{*0}}^2)}{(m_{B^{0}}+m_{K^{*0}})} (\varepsilon_{1}^{*} \cdot
p)(\varepsilon_{2}^{*} \cdot p) -i\frac {2 V^{B K^*}
 (m_{D^{*0}}^2)}{(m_{B^{0}}+m_{K^{*0}})}
\epsilon_{\mu \nu \alpha \beta} \varepsilon_{2}^{*\mu}
\varepsilon_{1}^{*\nu} p^{\alpha} p_1^{\beta}\biggr]\;,\label{amp}
\end{eqnarray}
where $f_{D^{*0}}$ is the decay constant
of the vector meson $D^{*0}$ and $\lambda_u^*=V_{ub}^* V_{cs}$.
Furthermore,
$A_1^{B K^*} (m_{D^{*0}}^2), A_2^{B K^*} (m_{D^{*0}}^2)$
and $V^{BK^*} (m_{D^{*0}}^2)$ are the form factors
 involved in the transition $B^0 \to K^{*0}$.
The coefficients $a_2$ is  given by
$a_{2} = C_2 + C_1/N_C$,
with $N_{C}$ as the number of colors.
 Thus, in this way, we can have the invariant amplitudes $a$, $b$ and $c$
(in the unit of $G_F/\sqrt 2 $) as
\begin{eqnarray}
a &=& i a_2 \lambda_u^*~f_{D^{*0}}m_{D^{*0}} (m_{B^{0}}+m_{K^{*0}})
A_1^{B K^*} (m_{D^{*0}}^2)\;,
\nonumber \\
b &=& -ia_2 \lambda_u^*~ f_{D^{*0}}m_{D^{*0}}\frac{2 m_{D^{*0}}
m_{K^{*0}}}{(m_{B^{0}}+m_{K^{*0}})} A_2^{B K^*} (m_{D^{*0}}^2)\;,
 \nonumber \\
c &=& -i a_2 \lambda_u^*~f_{D^{*0}} m_{D^{*0}} \frac{2 m_{D^{*0}}
m_{K^{*0}}}{(m_{B^{0}}+m_{K^{*0}})} V^{B K^*}(m_{D^{*0}}^2)
 \;.
\end{eqnarray}

Substituting the values of the effective coefficient $a_2 = 0.23$, the
 Wolfenstein parameters $A = 0.801, \lambda = 0.2265,
\bar \rho = 0.189$  and $\bar \eta = 0.358$ from \cite{charles},
the decay constant $f_{D^{*0}} = 240$ MeV,
the particle masses and lifetimes from \cite{pdg} and the form factors
$A_1^{B K^*} (m_{D^{*0}}^2) = 0.32,
A_2^{B K^*} (m_{D^{*0}}^2) = 0.31$ and $V^{B K^*}
(m_{D^{*0}}^2) = 0.52$ from \cite{ball},
we obtain the the branching ratio for the $B^0
\to D^{*0} K^{*0}$ as
\begin{eqnarray}
Br(B^0 \to D^{*0} K^{*0}) &=& 3.87 \times 10^{-6} \;.
\end{eqnarray}
Similarly, one can obtain the transition amplitude for the
 $\bar B^0 \to D^{*0} \bar K^{*0}$ process, which is analogous to
(\ref{amp}) with the replacement of $\lambda_u^*$ by $\lambda_c
=V_{cb} V_{us}^*$
and hence  the corresponding branching ratio as
\begin{eqnarray}
Br(\bar B^0 \to D^{*0} \bar K^{*0}) &=& 2.3 \times 10^{-5}\;.
\end{eqnarray}
Since the branching ratios of the above two processes are
 very much within the reach of the present experiments, we expect that
these processes will be measured soon by the currently running $B$ factories
and one will have a plenty of such events in the upcoming LHCb
factory.

Now, we consider the extraction of $(2\beta + \gamma)$ from the
modes  $\arp1~ \to \rp1~ ~\drp2~$. Since it is possible to obtain
the different helicity contributions by performing an angular
analysis \cite{dighe, matias}, from now onward we will concentrate
on the longitudinal (i.e., $A_0$) component, which is the dominant
one. The $K_S \pi^0$ mode of $\drp2~$ allows the $B^0 \to D^{*0}
K^{*0}$ and $\bar B^0 \to D^{*0} \bar K^{*0}$ amplitudes to
interfere with each other. As discussed earlier, the  decay
amplitude
 for the mode $B_{d}^0 \to D^{*0} K^{*0} $
arises from $\bar b \to \bar u c \bar s$ and  carries the weak phase
$e^{ i \gamma}$ while  $\bar B_{d}^0 \to D^{*0} \bar K^{*0}$ arises
from the quark transition $b \to c \bar u s$ and carries no weak
phase. The amplitudes also carry strong phases $e^{i \delta _1}$ and
$e^{i \delta _2}$. Thus, we can write the longitudinal
components of the decay amplitudes as
\begin{eqnarray}
A_{0}(f) &=& {\rm Amp} {(B_{d}^0 \to f)_{0}} = M_1 e^{
i \gamma} e^{i \delta _1} \nonumber \;, \\
\bar A_{0}(f) &=& {\rm Amp }{(\bar B_{d}^0 \to f)_{0}} =
M_2 e^{i \delta_2} \nonumber \;, \\
\bar A_{0}(\bar f ) &=& {\rm Amp} {(\bar B_{d}^0 \to \bar f)_{0}} =
M_1 e^{-i \gamma} e^{i \delta _1} \nonumber \;,
\\
 A_{0}(\bar f) &=& {\rm Amp} {(B_{d}^0 \to \bar
f)_{0}} = M_2 e^{i \delta_2}.\label{sd}
\end{eqnarray}

Since the final state  $f = D^{*0} \drp2 ~$ is  accessible to $B^0$
and  $\bar B^0$, inserting the time evolution of the observables
$A_0(t)$ as in \cite{time},  one arrives at the usual expression for
the longitudinal component of the time dependent decay widths
\cite{bigi} as
\begin{eqnarray}
\Gamma_0 (B^0(t) \to f) &=& \frac{ e^{-\Gamma t}}{2}~
 \biggr\{ \left (|A_0(f)|^2+|\bar A_0(f)|^2 \right )
 +\left (|A_0(f)|^2-|\bar A_0(f)|^2 \right )\cos \Delta m t \nonumber\\
&-&2{\rm Im} \left (\frac{q}{p}A_0(f)^* \bar A_0(f)
\right )
\sin \Delta m t\biggr\},
\nonumber\\
 \Gamma_0 (\bar B^0(t) \to f)&=& \frac{e^{-\Gamma t}}{2}~
 \biggr\{\left (|A_0(f)|^2+|\bar A_0(f)|^2\right )
-\left [|A_0(f)|^2-|\bar A_0(f)|^2\right ) \cos \Delta m t \nonumber\\
&+&2  {\rm Im}
\left (\frac{q}{p}A_0(f)^* \bar A_0(f)\right )
 \sin \Delta m t\biggr\},\label{dl2}
\end{eqnarray}
where $q/p={\rm exp}(-2 i\beta)$ is the $B^0 - \bar B^0$ mixing
parameter and $\Gamma $ and $\Delta m$  denote the average width and
the mass difference of the heavy and light $B$ mesons and we have
neglected the small width difference $\Delta \Gamma$ between them.

Thus, the time dependent measurement of
the longitudinal component of  $B^0(t)
\to f $ decay rates allow one to obtain
the following observables :
\begin{equation}
|A_0(f)|^2+|\bar A_0(f)|^2,~~~~ |A_0(f)|^2-|\bar A_0(f)|^2,
~~{\rm and}~~{\rm Im}[\frac{q}{p}A_0(f)^* \bar A_0(f)],
\end{equation}
i.e., the longitudinal components of CP averaged branching ratio,
the direct CP violation and the mixing induced CP violation
parameters.

Similarly one can obtain the time dependent decay rates for the
final state $\bar f$ i.e., $\Gamma_0(\bar B^0(t) \to \bar f)$ from
by $\Gamma_0( B^0(t) \to  f)$ by replacing $A_0(f)$ by $\bar
A_0(\bar f)$ and $\bar A_0(f)$ by corresponding CP conjugate $
A_0(\bar f)$. $\Gamma_0(B^0(t) \to \bar f)$ can be obtained from
$\Gamma_0(\bar B^0(t) \to f)$ with similar substitution.

 Now substituting the decay amplitudes as
defined in Eq. (\ref{sd}) in (\ref{dl2}) we get the decay rates as
\begin{eqnarray}
\Gamma_0 (B^0(t)   \to  f) & = &\frac{ e^{-\Gamma t}}{2} ~
 \biggr\{ (M_1^2+M_2^2)
+(M_1^2- M_2^2 )\cos \Delta m t \nonumber\\
&-& 2  M_1  M_2 \sin (\delta -\phi)
\sin \Delta m t\biggr\}\;,
\nonumber\\
\Gamma_0 (B^0(t)  \to  \bar f)&=& \frac{ e^{-\Gamma t}}{2}~
\biggr\{(M_1^2+ M_2^2) -(M_1^2- M_2^2) \cos \Delta m t
 \nonumber\\
& +&  2  M_1  M_2 \sin (\delta +\phi)
 \sin \Delta m t\biggr\}\;,
\nonumber\\
\Gamma_0 (\bar B^0(t)   \to  \bar f)&=& \frac { e^{-\Gamma t}}{2}~
\biggr\{ (M_1^2+M_2^2) + (M_1^2-M_2^2)\cos \Delta m t
\nonumber\\
&-&2 M_1  M_2 \sin (\delta +\phi)
 \sin \Delta m t\biggr\}\;,
\nonumber\\
 \Gamma_0 (\bar B^0(t) \to  f) &=& \frac{e^{-\Gamma t}}{2}~
 \biggr\{(M_1^2+M_2^2)
-(M_1^2-M_2^2) \cos \Delta m t \nonumber\\
&+&2  M_1  M_2 \sin (\delta -\phi)
 \sin \Delta m t\biggr\}\;,
\end{eqnarray}
where $\delta = \delta_2-\delta_1$ is the strong phase difference
between the longitudinal components of the two amplitudes
$\bar{B}^0\to f$ and $B^0 \to f $ and $\phi= 2\beta+\gamma$. Thus
through the measurements of the time dependent rates, it is possible
to measure the amplitudes $M_1$ and $M_2$ and the CP violating
quantities $S_+ \equiv \sin(\delta+\phi)$ and $S_- \equiv \sin
(\delta - \phi) $. In turn these quantities will determine
$\tan^2\phi $ up to a four fold ambiguity via the expression
\begin{equation}
\tan^2 \phi [\cot^2 \delta] = \frac{(S_+ - S_-)^2}{2 - S_-^2 - S_+^2
\pm 2 \sqrt{(1 - S_+^2) (1 - S_-^2)}}\;,
\end{equation}
where one sign will give $\tan^2 \phi $
 and the other $\cot^2 \delta$.

Let us now estimate the number of reconstructed events that could be
observed at the B factories assuming that $3 \times 10^8$
($10^{12}$) $B \bar B$'s are (will be) available at the $e^+ e^-$
asymmetric $B$ factories (hadronic $B$ machines like LHCb). Let us
first estimate the number of $B^0 \to D^0 (\bar D^0) K^{*0} $ events
will be available in the upcoming LHCb experiment. Assuming the
branching ratio for the process $B^0 \to D^0 K^{*0}$ to be $|(V_{ub}
V_{cs}^*)/(V_{cb}V_{us}^*)|^2 \times Br(B^0 \to \bar D^0 K^{*0})
\approx 8.5 \times 10^{-6}$, $Br(D^0 \to K^{*+}K^-)=3.7 \times
10^{-3}$ \cite{pdg} and $10 \%$ overall reconstruction efficiency,
we expect to get nearly $ 9 \times 10^3 $ events per year of running
at LHCb. For the corresponding vector-vector modes, we use the
longitudinal component of the branching ratio as $Br_0(B^0 \to
D^{*0} K^{*0})=0.65\times Br(B^0 \to D^{*0} K^{*0}) \approx 2.51
\times 10^{-6} $, $Br(D^{*0} \to D^0 \pi^0)=62 \%$ \cite{pdg},
$Br(K^{*0} \to K_S \pi^0)=Br(K^{*0} \to K \pi)/3$, and an overall
efficiency of $10 \%$. Thus we expect to get approximately 15 ($5
\times 10^4$) reconstructed events at the $e^+ e^-$ (hadronic)
machines per year of running. This crude estimate indicates that
this method may be well suited for the the extraction of the weak
phase $\gamma (2 \beta+ \gamma)$ at LHCb.

In this paper, we have carried out a study of the color suppressed
decay modes $B^0 \to D^0 (D^{0 *}) K^{0*}$ to extract the weak phase
$\gamma (2 \beta + \gamma)$. For the extraction of $\gamma$ we
considered the decay modes  $B^0 \to D^0 (\bar D^0)  K^{0*}$, with
subsequent decay of $D^0(\bar D^0)$ into non-CP state $K^{*+}K^-$.
The use of the non-CP state allows the two interfering amplitudes to
be of same order and hence one can cleanly extract the CKM angle
$\gamma$. Next we considered the processes $B^0 \to D^{*0} (\bar
D^{*0})  K^{0*}$, where the final states are admixtures of CP-even
and CP-odd states. However, it is possible to disentangle them using
the angular distributions of the final decay products. Now
considering the longitudinal component of the time-dependent decay
rates of these modes  we have shown that $\phi \equiv (2 \beta
+\gamma)$ can be cleanly obtained. Since these modes are free from
penguin pollution and also the branching ratios are measurable at
hadron factories such as the LHCb, we feel that they could be very
much suited for determining the phase $\gamma (2\beta + \gamma)$.

\acknowledgments
The work of RM was partly supported by Department of Science and Technology,
Government of India, through grant No.
SR/S2/HEP-04/2005. AKG and BM would like to thank Council of Scientific
and Industrial Research, Government of India, for financial support.

\end{document}